\documentclass[runningheads]{llncs}

\usepackage[T1]{fontenc}
\usepackage[utf8]{inputenc}
\usepackage{amsmath,amssymb,amsfonts}
\usepackage{booktabs}
\usepackage{tabularx}
\usepackage{array}
\usepackage{multirow}
\usepackage{graphicx}
\usepackage{xcolor}
\usepackage{hyperref}
\usepackage{url}
\usepackage{enumitem}
\usepackage{tikz}
\usepackage{pgfplots}
\usepackage{listings}
\usepackage{float}
\pgfplotsset{compat=1.18}

\usetikzlibrary{arrows.meta,positioning,shapes.geometric,fit,calc}

\hypersetup{
  colorlinks=true,
  linkcolor=blue!55!black,
  citecolor=blue!55!black,
  urlcolor=blue!55!black
}

\setlist[itemize]{leftmargin=1.2em,itemsep=0.25em,topsep=0.25em}
\setlist[enumerate]{leftmargin=1.4em,itemsep=0.25em,topsep=0.25em}

\lstdefinestyle{terminal}{
  basicstyle=\ttfamily\footnotesize,
  breaklines=true,
  columns=fullflexible,
  keepspaces=true,
  frame=single,
  rulecolor=\color{black!35},
  backgroundcolor=\color{black!2}
}

\newcommand{\qb}{\textsc{QBalance}}
\newcommand{\pkg}{\texttt{qbalance}}

\newcommand{\R}{\mathbb{R}}

\title{\qb: A Reproducible Multi-Objective Workflow for Quantum Compilation, Noise Suppression, and Error-Mitigation Strategy Selection}

\titlerunning{\qb{} for Quantum Compilation and Mitigation Strategy Selection}

\author{Soumyadip Sarkar}
\authorrunning{S. Sarkar}
\institute{Independent Researcher}

\begin{document}
\maketitle

\begin{abstract}
Near-term quantum workloads are shaped by coupled compilation and execution choices: qubit layout, routing, basis translation, gate suppression, measurement mitigation, shot budget, and artifact reproducibility. This paper analyzes \qb{}, a Python workflow library for dataset-level selection among quantum compilation, noise-suppression, and error-mitigation strategies built on the Qiskit ecosystem. The contribution is formulated as a finite multi-objective strategy-selection problem over circuits, backends, and transformation policies. The manuscript derives the implemented weighted objective, non-dominated selection rule, survival-product error proxy, Bayesian linear candidate-ordering surrogate, and distributional diagnostics. It also positions the system relative to established work on Qiskit pass-manager compilation, SABRE-style routing, randomized compiling, dynamical decoupling, zero-noise extrapolation, matrix-free measurement mitigation, circuit cutting, and Thompson sampling. The analysis shows that \qb{} provides a reproducible orchestration and artifact model for quantum workflow studies. It also establishes precise limitations: the current bandit mechanism orders candidates but does not reduce the number of candidate evaluations, the custom layout heuristic is greedy and only partially topology-aware, the implemented ZNE helper is parity-centered, and the cutting integration is a hook rather than a full reconstruction pipeline.
\let\thefootnote\relax\footnote{The code can be found at: \url{https://github.com/neuralsorcerer/qbalance}}
\end{abstract}

\section{Introduction}

Noisy intermediate-scale quantum devices make compilation and execution a coupled optimization problem. A circuit that is shallow in an abstract gate set may become deep after basis translation and routing. A routing choice that reduces swaps may still be poor if it places the most active logical qubits on low-quality physical qubits. A mitigation strategy may improve an estimator but increase sampling cost, classical post-processing cost, or execution latency. As a result, a practical workflow cannot be evaluated using a single compilation flag or a single circuit instance.

\qb{} addresses this setting as a workflow library for balancing compilation, suppression, and mitigation choices over datasets of quantum circuits. The analyzed implementation is version 0.1.0, requires Python $\geq 3.10$, depends on Qiskit $\geq 2.0$, and exposes optional integrations for Qiskit Aer, Qiskit Runtime, M3, circuit cutting, and reporting \cite{qbalanceRepo}. The implementation uses Qiskit's preset pass-manager mechanism for target-aware transpilation, a documented route for constructing staged pass managers and applying them to circuits \cite{QiskitPassManagers2026,JavadiAbhari2024Qiskit}. Optional count execution is backend-compatible and can use Qiskit Aer when a backend object is not directly runnable \cite{QiskitAerSimulator2026}.

The central question is the comparison of a finite set of candidate strategies across a circuit dataset when each strategy changes several coupled quantities. The paper therefore treats \qb{} as a deterministic, inspectable selection system rather than as an empirical claim about hardware advantage. All conclusions are restricted to properties supported by the inspected implementation and by cited literature. Because the project does not bundle a large benchmark corpus, hardware calibration snapshots, or complete experimental result matrices, this manuscript does not claim device-level accuracy improvements or reductions in physical error rates.

\subsection{Problem Statement}

The paper answers four questions.

\begin{description}[leftmargin=1.7em,style=nextline]
\item[Q1.] What optimization problem is solved by \qb{} at the dataset-workflow level?
\item[Q2.] How do the implemented objective, Pareto selector, and Bayesian linear search surrogate behave mathematically?
\item[Q3.] Which parts of the system are direct orchestration over established Qiskit and mitigation mechanisms, and which parts are repository-specific heuristics?
\item[Q4.] What correctness, reproducibility, and complexity properties can be concluded from the released source code without inventing experimental results?
\end{description}

\subsection{Contributions}

This study makes the following contributions.

\begin{enumerate}
\item It formalizes \qb{} as a dataset-level multi-objective strategy-selection workflow over circuits, backends, and strategy specifications.
\item It derives the implemented scalar objective, non-dominated selection rule, error-survival proxy, Bayesian linear posterior, and distributional diagnostics.
\item It consolidates software and literature references into a compact bibliography, citing the repository as the implementation artifact and using peer-reviewed or official sources for technical mechanisms.
\item It states analytical results about the default candidate space, candidate-evaluation complexity, finite-safe scoring, and the current limits of the bandit, zero-noise extrapolation, measurement twirling, and circuit-cutting paths.
\item It supplies a reproducibility protocol and an artifact checklist suitable for later empirical evaluation.
\end{enumerate}

\section{Background and Related Work}

\subsection{Qiskit Compilation and Pass Managers}

Qiskit represents quantum programs as circuits and compiles them through staged transformations such as layout selection, routing, basis translation, optimization, and scheduling \cite{JavadiAbhari2024Qiskit}. Routing is especially important on NISQ devices because limited connectivity can require additional SWAP operations; SABRE is a well-known bidirectional heuristic for this mapping problem \cite{Li2019SABRE}. The official documentation presents preset pass managers as a mechanism for constructing a pass manager with reasonable defaults. A preset pass manager can be configured by optimization level, backend, target, basis gates, coupling map, layout method, routing method, translation method, scheduling method, and transpiler seed \cite{QiskitPassManagers2026}. \qb{} uses this mechanism as the compilation substrate rather than implementing a new low-level transpiler.

\subsection{Noise Suppression}

Randomized compiling and Pauli twirling aim to transform coherent or structured errors into more stochastic effective noise while preserving the logical circuit in expectation. Wallman and Emerson formalized randomized compiling as a way to tailor noise for scalable quantum computation \cite{Wallman2016}. If $\mathcal{E}$ is a noisy operation and $\mathcal{P}$ is a twirling group, the twirled channel has the schematic form
\begin{equation}
\mathcal{E}_{\mathrm{twirl}}(\rho)
  = \frac{1}{|\mathcal{P}|}\sum_{P \in \mathcal{P}} P^{\dagger}\,\mathcal{E}(P\rho P^{\dagger})\,P,
\label{eq:twirled-channel}
\end{equation}
with the exact circuit identity depending on gate set and correction placement. \qb{} delegates two-qubit Pauli-twirling circuit generation to Qiskit's circuit-level functionality and then compiles the resulting ensemble \cite{qbalanceRepo}.

Dynamical decoupling inserts carefully chosen pulse or gate sequences into idle periods. In a first-order toggling-frame view, a control sequence $\{g_j\}_{j=1}^{L}$ transforms an unwanted system-environment Hamiltonian $H$ into an average Hamiltonian
\begin{equation}
\overline{H}^{(0)} = \frac{1}{L}\sum_{j=1}^{L} g_j^{\dagger} H g_j,
\end{equation}
which can suppress selected error terms when the group action averages them toward zero. Viola, Knill, and Lloyd introduced dynamical decoupling as a method for filtering unwanted interactions in open quantum systems \cite{Viola1999}. \qb{} builds a Qiskit pass manager using ALAP scheduling and dynamical-decoupling padding when the strategy enables this knob \cite{qbalanceRepo,QiskitPassManagers2026}.

\subsection{Error Mitigation}

Zero-noise extrapolation (ZNE) estimates a zero-noise observable by evaluating the computation at amplified noise levels and extrapolating the observable to noise scale zero. Temme, Bravyi, and Gambetta introduced this idea for short-depth quantum circuits \cite{Temme2017}, and error mitigation more broadly has become a central technique for near-term applications \cite{Endo2018}. If $x_i\geq 1$ are noise-scale factors and $y_i$ are observed expectation values, polynomial ZNE fits
\begin{equation}
 y_i = \sum_{r=0}^{d} a_r x_i^r + \varepsilon_i,
 \qquad \widehat{y}(0)=\widehat{a}_0.
\label{eq:zne-fit}
\end{equation}
\qb{} implements a compact counts-based ZNE helper that extrapolates a parity expectation and returns a pseudo-probability distribution adjusted to match the extrapolated parity mass. This is narrower than a general observable-level ZNE interface, and the paper treats it as a repository-specific heuristic rather than a complete mitigation framework.

Measurement mitigation corrects readout noise after measurement and is part of the broader family of near-term quantum error mitigation techniques surveyed in recent reviews \cite{Cai2023QEM}. M3 is a matrix-free measurement mitigation approach that works in a reduced subspace defined by observed noisy bit strings, avoiding explicit construction and inversion of the full assignment matrix \cite{Nation2021}. The M3 documentation describes it as a matrix-free routine whose reduced linear system can be much smaller than the full Hilbert-space assignment matrix \cite{MthreeDocs2026}. \qb{} invokes M3 through its optional mitigation path and records summary probabilities rather than replacing the complete workflow output with a full mitigation study.

\subsection{Circuit Cutting}

Circuit cutting decomposes a circuit into smaller subexperiments that can be run on smaller devices or partitions, at the cost of increased sampling overhead and reconstruction work. CutQC showed how circuit cutting can use smaller quantum computers for larger circuit evaluations \cite{Tang2021CutQC}. The Qiskit addon documentation states that cutting creates smaller circuits and reconstructs the original result through classical post-processing, while the total number of shots must increase by a factor determined by the cuts, usually called sampling overhead \cite{QiskitAddonCutting2026}. \qb{} includes an optional cut-finding hook but does not implement the full reconstruction pipeline end to end.

\subsection{Bandit Search and Multi-Objective Selection}

The \qb{} search routine uses a Bayesian linear model to rank strategies, which is naturally related to linear Thompson sampling. Agrawal and Goyal studied Thompson sampling for contextual bandits with linear payoffs and provided theoretical guarantees for the contextual setting \cite{Agrawal2013}. \qb{} uses a simpler model: the feature vector describes only the strategy, not the circuit. Therefore the current implementation is best described as a Thompson-style linear ordering surrogate, not a full contextual bandit over circuit and strategy pairs.

For distribution diagnostics, \qb{} computes Kolmogorov-Smirnov, Cramer-von Mises type, and earth-mover distances on one-dimensional metric samples. These tests and distances are standard tools for comparing empirical distributions; Darling's treatment of the Kolmogorov-Smirnov and Cramer-von Mises tests is a canonical statistical reference \cite{Darling1957}.

\section{Problem Formulation}

Let
\begin{equation}
\mathcal{D}=\{C_1,C_2,\ldots,C_N\}
\end{equation}
be a dataset of quantum circuits. Let $B$ denote a target backend or backend-like simulator, and let $\mathcal{S}$ denote a finite set of strategy specifications. A strategy $s\in\mathcal{S}$ contains compilation knobs, suppression flags, mitigation flags, and optional cutting metadata. For a circuit $C_i$, backend $B$, and strategy $s$, define
\begin{equation}
T(C_i,B,s) \rightarrow (\widetilde{C}_{i,s}, m_{i,s}),
\end{equation}
where $\widetilde{C}_{i,s}$ is the transformed circuit and $m_{i,s}$ is a metric dictionary containing measurable or estimated quantities such as depth, size, width, two-qubit operation count, estimated error, compile time, execution entropy, top observed probability, and mitigation summaries.

\subsection{Scalar Objective}

The default scalar objective is a weighted sum over available finite metrics:
\begin{equation}
J_w(m)=\sum_{k\in K}w_k m_k.
\label{eq:objective-general}
\end{equation}
The implemented default weights are
\begin{equation}
J_w(m)=1.0\,m_{\mathrm{depth} }
+2.0\,m_{\mathrm{2q}}
+10.0\,m_{\mathrm{err}}
+0.1\,m_{\mathrm{time}}.
\label{eq:objective-default}
\end{equation}
Here $m_{\mathrm{2q}}$ is the number of two-qubit operations, $m_{\mathrm{err}}$ is the internal estimated-error proxy, and $m_{\mathrm{time}}$ is compile time in seconds. Lower values are preferred. The implementation ignores missing, non-numeric, and non-finite terms during scoring, and its final fallback logic avoids giving malformed candidates an accidental zero score \cite{qbalanceRepo}.

\subsection{Pareto Selection}

When Pareto selection is enabled, \qb{} first filters candidates using the metric tuple
\begin{equation}
\psi(m)=\left(m_{\mathrm{depth}},m_{\mathrm{2q}},m_{\mathrm{err}}\right).
\end{equation}
For candidates $a$ and $b$, candidate $a$ dominates $b$ when
\begin{equation}
\left(\forall r,\; \psi_r(a)\leq \psi_r(b)\right)
\wedge
\left(\exists r,\; \psi_r(a)<\psi_r(b)\right).
\label{eq:dominance}
\end{equation}
The Pareto front is
\begin{equation}
\mathcal{P}=\{s\in\mathcal{S}:\nexists s'\in\mathcal{S}\;\text{such that}\;s'\prec s\},
\end{equation}
where $s'\prec s$ denotes dominance under Eq.~\eqref{eq:dominance}. The selected strategy is
\begin{equation}
s_i^* = \arg\min_{s\in\mathcal{P}} J_w(m_{i,s}).
\end{equation}
If Pareto filtering is disabled, the selected strategy is the minimum of $J_w$ over all successfully evaluated candidates.

\subsection{Compilation and Execution Objective Boundary}

It is important to separate compile-time metrics from execution-time metrics. In compile-only mode, \qb{} can score depth, two-qubit count, estimated error, and compile time. In execution-enabled mode, counts-dependent diagnostics and mitigation summaries can also be recorded. The implementation does not make the default objective depend on raw-count entropy, top observed probability, M3 output, or ZNE output. Those quantities are available as additional metrics, but the default objective remains the four-term scalarization in Eq.~\eqref{eq:objective-default}.

\section{System Design}

\subsection{Workflow Overview}

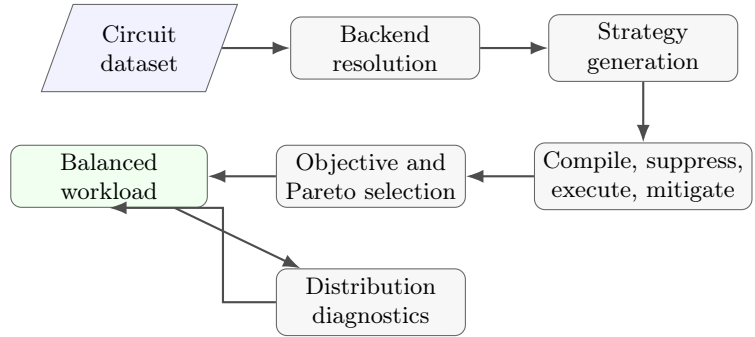
\begin{figure}[t]
\centering
\begin{tikzpicture}[
  node distance=8mm and 9mm,
  box/.style={rectangle,rounded corners,draw=black!60,fill=black!3,align=center,minimum width=25mm,minimum height=8mm,font=\small},
  data/.style={trapezium,trapezium left angle=70,trapezium right angle=110,draw=black!60,fill=blue!5,align=center,minimum width=26mm,minimum height=8mm,font=\small},
  outbox/.style={rectangle,rounded corners,draw=black!60,fill=green!6,align=center,minimum width=26mm,minimum height=8mm,font=\small},
  arr/.style={-Latex,thick,draw=black!70}
]
\node[data] (ds) {Circuit\\dataset};
\node[box,right=of ds] (backend) {Backend\\resolution};
\node[box,right=of backend] (cand) {Strategy\\generation};
\node[box,below=of cand] (eval) {Compile, suppress,\\execute, mitigate};
\node[box,left=of eval] (score) {Objective and\\Pareto selection};
\node[outbox,left=of score] (out) {Balanced\\workload};
\node[box,below=of score] (diag) {Distribution\\diagnostics};
\draw[arr] (ds) -- (backend);
\draw[arr] (backend) -- (cand);
\draw[arr] (cand) -- (eval);
\draw[arr] (eval) -- (score);
\draw[arr] (score) -- (out);
\draw[arr] (out) -- (diag);
\draw[arr] (diag.west) -- ++(-7mm,0) |- (out.south);
\end{tikzpicture}
\caption{Dataset-level \qb{} workflow. The system is an orchestration layer over compilation, suppression, execution, mitigation, scoring, and artifact persistence rather than a replacement for Qiskit's low-level compiler.}
\label{fig:workflow}
\end{figure}

Figure~\ref{fig:workflow} presents the dataset-level workflow. A dataset is loaded, a backend is resolved, a candidate strategy set is constructed, a baseline is compiled, strategies are evaluated, and one strategy is selected for each circuit. The output is a balanced workload with persisted metadata, copied dataset artifacts, and summary diagnostics.

\subsection{Strategy Specification}

A strategy specification is an immutable object with five groups of knobs:

\begin{enumerate}
\item \textbf{Compilation}: optimization level, layout method, routing method, translation method, and transpiler seed.
\item \textbf{Suppression}: Pauli twirling, number of twirls, dynamical decoupling, decoupling sequence, measurement twirling, and suppression seed.
\item \textbf{Mitigation}: M3 flag, ZNE flag, ZNE scale factors, and polynomial degree.
\item \textbf{Circuit cutting}: cutting flag and maximum subcircuit qubit constraint.
\item \textbf{Runtime metadata}: an optional resilience-level field.
\end{enumerate}

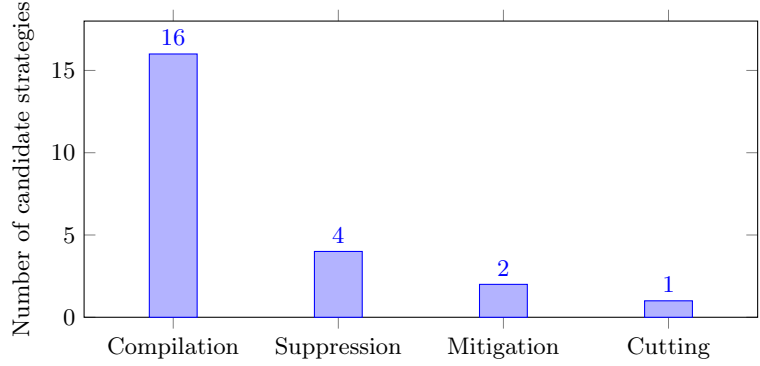
\begin{figure}[t]
\centering
\begin{tikzpicture}
\begin{axis}[
  ybar,
  width=0.86\textwidth,
  height=5.5cm,
  ymin=0,
  ymax=18,
  ylabel={Number of candidate strategies},
  symbolic x coords={Compilation,Suppression,Mitigation,Cutting},
  xtick=data,
  nodes near coords,
  bar width=18pt,
  enlarge x limits=0.18
]
\addplot coordinates {(Compilation,16) (Suppression,4) (Mitigation,2) (Cutting,1)};
\end{axis}
\end{tikzpicture}
\caption{Default candidate-space composition in \qb{} version 0.1.0. The total is 23, although user-facing examples commonly set a maximum of 24.}
\label{fig:candidate-count}
\end{figure}

The default candidate generator produces a compact finite set. The first block sweeps optimization levels $\{0,1,2,3\}$ with four compilation variants per level. It then adds suppression, mitigation, and cutting variants. Because duplicates are removed and the default maximum is 24, the released generator yields at most 23 unique default strategies. Figure~\ref{fig:candidate-count} shows the composition.

\subsection{Dataset and Artifact Model}

The dataset model consists of a JSON index plus serialized circuit artifacts. The implementation supports QPY and QASM loading, saves new datasets as QPY, validates safe single-component artifact paths, and records optional metadata. This design is important for reproducibility because it separates circuit identity from local Python objects and gives every workflow run a portable dataset directory. The built-in tiny dataset contains three circuits: a Bell circuit, a three-qubit GHZ circuit, and a four-qubit QFT-like circuit. These examples are suitable for smoke tests and workflow validation, but they are too small to support broad performance claims.

\subsection{Backend and Optional Dependency Model}

\qb{} resolves backend strings through plugin entry points. The package declares built-in backend entry points for fake and Aer backends. Optional extras control whether Aer simulation, Runtime helper support, M3, circuit cutting, and reporting dependencies are installed. This keeps the base package lightweight while allowing heavier integrations when needed. The Qiskit Runtime documentation describes primitive options as grouped settings, with common top-level options such as resilience level and nested categories such as execution, resilience, simulator, and twirling \cite{QiskitRuntimeOptions2026}. \qb{} currently provides a helper for constructing Runtime-style option dictionaries, but the main workflow does not execute through Runtime primitives.

\section{Methods}

\subsection{Compilation Procedure}

For each candidate strategy, \qb{} constructs a compilation path around Qiskit's preset pass managers. In abstract form, the compilation procedure is
\begin{align}
C &\xrightarrow{\text{optional twirling}} \{C^{(1)},\ldots,C^{(T)}\}\notag\\
  &\xrightarrow{\text{preset pass manager}} \{\widetilde{C}^{(1)},\ldots,\widetilde{C}^{(T)}\}\notag\\
  &\xrightarrow{\text{optional DD and measurement twirling}} \text{candidate output}.
\end{align}
When Pauli twirling produces an ensemble of size $T$, \qb{} compiles each member and selects the member minimizing the internal proxy
\begin{equation}
S_{\mathrm{ensemble}}(m)=m_{\mathrm{depth}}+10m_{\mathrm{err}}.
\label{eq:ensemble-proxy}
\end{equation}
This proxy is not identical to the default outer objective because it excludes two-qubit operation count and compile time. Therefore twirl-ensemble selection and final strategy selection are distinct optimization layers.

\subsection{Estimated Error Proxy}

The implementation estimates circuit error using a survival-product approximation. Let $e_1,\ldots,e_L$ be operation-level error proxies inferred from backend calibration data when possible and conservative fallback constants otherwise. The survival probability is approximated by
\begin{equation}
P_{\mathrm{survive}}=\prod_{\ell=1}^{L}(1-e_{\ell}),
\end{equation}
and the estimated error is
\begin{equation}
\widehat{E}=1-P_{\mathrm{survive}}=1-\prod_{\ell=1}^{L}(1-e_{\ell}).
\label{eq:survival-error}
\end{equation}
For small $e_{\ell}$,
\begin{equation}
\widehat{E}=\sum_{\ell=1}^{L}e_{\ell}-\sum_{\ell<r}e_{\ell}e_r+O(e^3)\approx\sum_{\ell=1}^{L}e_{\ell}.
\end{equation}
Thus Eq.~\eqref{eq:survival-error} is a reasonable coarse proxy for error exposure, but it is not a calibrated physical noise model. It ignores coherent error structure, crosstalk, temporal drift, pulse-level scheduling details, and observable-specific sensitivity.

\subsection{Noise-Aware Layout Heuristic}

The custom layout mode ranks logical qubits by two-qubit interaction activity and physical qubits by a single-qubit quality score. If $\deg(q)$ is the number of two-qubit interactions involving logical qubit $q$, the logical order sorts $q$ by decreasing $\deg(q)$. For physical qubit $p$, the implemented quality score is
\begin{equation}
Q(p)=(1-r_p)+10^{-5}\left(T_{1,p}+T_{2,p}\right),
\label{eq:quality-score}
\end{equation}
where $r_p$ is readout error and $T_{1,p},T_{2,p}$ are coherence-time fields when available. Missing values are replaced by fallback constants. The most active logical qubits are greedily assigned to the highest-scoring physical qubits.

This heuristic is simple and inexpensive. It is also intentionally limited. It does not solve a subgraph embedding problem, does not directly minimize routing distance, and does not explicitly optimize pairwise two-qubit gate error. A topology-aware extension would score pairs $(p_i,p_j)$ and include coupling-map distance or calibrated two-qubit error in the placement objective.

\subsection{Bayesian Linear Ordering Surrogate}

In bandit mode, \qb{} featurizes each strategy $s$ into a fixed vector $\phi(s)\in\R^{12}$. The features include a bias term, optimization level, routing and layout indicators, suppression indicators, mitigation indicators, and the cutting flag. After observing scores $y_1,\ldots,y_t$ for strategies $s_1,\ldots,s_t$, the model assumes
\begin{equation}
 y = Xw+\varepsilon,\qquad \varepsilon\sim\mathcal{N}(0,\sigma^2 I),\qquad w\sim\mathcal{N}(0,\alpha^{-1}I),
\end{equation}
where $X$ has rows $\phi(s_j)^\top$. The posterior precision and mean are
\begin{align}
\Lambda &= \alpha I + \frac{1}{\sigma^2}X^\top X,\label{eq:posterior-precision}\\
\mu &= \Lambda^{-1}\frac{1}{\sigma^2}X^\top y.\label{eq:posterior-mean}
\end{align}
To sample without materializing the covariance, the implementation computes a Cholesky factorization $\Lambda=LL^\top$ and draws
\begin{equation}
\widetilde{w}=\mu+L^{-\top}z,\qquad z\sim\mathcal{N}(0,I).
\end{equation}
Since
\begin{equation}
\mathrm{Cov}(L^{-\top}z)=L^{-\top}L^{-1}=\Lambda^{-1},
\end{equation}
this has the intended Gaussian posterior covariance. Candidate strategies are ordered by the sampled score $\phi(s)^\top\widetilde{w}$, with lower values preferred. The method is Thompson-style because it samples a plausible linear model and acts greedily under that sampled model \cite{Agrawal2013}.

\paragraph{Important distinction.}
The current implementation ranks candidates but does not reduce the number of candidates evaluated. Let $C=|\mathcal{S}|$. Both grid and bandit modes attempt to evaluate the complete candidate order for each circuit, except for candidates that fail. Therefore the bandit path does not provide a lower asymptotic candidate-evaluation count in version 0.1.0. It can only change order and cross-circuit learning behavior.

\subsection{Distributional Diagnostics}

For a metric $k$, let $X=\{x_i\}_{i=1}^N$ be baseline values and $Y=\{y_i\}_{i=1}^N$ be selected-workflow values. Let $F_X$ and $F_Y$ be their empirical CDFs on an aligned support grid. \qb{} computes
\begin{align}
D_{\mathrm{KS}}(X,Y) &= \sup_z |F_X(z)-F_Y(z)|,\\
W_1(X,Y) &= \int |F_X(z)-F_Y(z)|\,dz,\\
D_{\mathrm{CVM}}(X,Y) &= \int (F_X(z)-F_Y(z))^2\,dz.
\end{align}
The first quantity is the Kolmogorov-Smirnov distance, the second is the one-dimensional Wasserstein-1 or earth-mover form, and the third is a Cramer-von Mises type squared-CDF distance \cite{Darling1957}. These diagnostics compare distributions of metrics across circuits rather than proving physical correctness.

\section{Algorithmic Specification}

Algorithm~\ref{alg:adjust} summarizes the dataset-level adjustment procedure.

\begin{figure}[ht]
\begin{lstlisting}[style=terminal,caption={Algorithmic view of dataset-level strategy adjustment.},label={alg:adjust}]
Input: dataset D, backend B, strategy set S, objective J, search mode M
Output: selected strategy s_i for each circuit C_i

1. Resolve the backend object from the backend specification.
2. Generate the finite candidate strategy set S.
3. Compile each circuit once with the fixed baseline strategy.
4. For each circuit C_i in D:
      a. Choose an evaluation order over S.
         - grid: use the default candidate order.
         - bandit: random warmup, then Thompson-style proposed order.
      b. For each strategy s in the chosen order:
            i. Optionally run cut finding.
           ii. Compile C_i under s using a preset pass manager.
          iii. Optionally execute the compiled circuit.
           iv. Optionally apply measurement untwirling, M3, or ZNE helpers.
            v. Compute J(m_i,s) and record all metrics.
           vi. If M is bandit, update the linear surrogate with the score.
      c. If Pareto mode is enabled, filter to non-dominated candidates.
      d. Select the candidate with minimum objective score.
5. Return a balanced workload containing selections and baseline metrics.
\end{lstlisting}
\end{figure}

\section{Correctness and Complexity Analysis}

\subsection{Finite-Safe Objective Scoring}

\textbf{Proposition 1.} Under the implemented fallback rule, a candidate whose metrics contain no finite objective-relevant term cannot be preferred merely because invalid values were skipped.

\textit{Argument.} The primary scoring function ignores missing, non-numeric, and non-finite values. That alone could make a completely invalid metric map produce score zero. The selection fallback addresses this by deriving a finite-safe objective from the raw metrics and returning $+\infty$ when objective keys are present but none contributes a finite term. Therefore invalid candidates are treated as worst-case during selection rather than as artificially optimal. This is a correctness property of the selector, not a statistical guarantee about metric quality.

\subsection{Candidate-Evaluation Complexity}

Let $N$ be the number of circuits, $C$ the number of constructed candidates, $T_s$ the number of twirled circuits constructed for strategy $s$, $R(C_i,B,s)$ the cost of one compilation, $E(C_i,B,s)$ the cost of one execution, and $F_s$ the number of ZNE scale factors for strategy $s$. Compile-only adjustment has dominant cost
\begin{equation}
O\left(NR_{\mathrm{base}}+\sum_{i=1}^{N}\sum_{s\in\mathcal{S}}T_sR(C_i,B,s)\right).
\end{equation}
When execution and ZNE are enabled, the cost becomes
\begin{align}
O\Big(&NR_{\mathrm{base}}+\sum_{i=1}^{N}\sum_{s\in\mathcal{S}}T_sR(C_i,B,s)\notag\\
&+\sum_{i=1}^{N}\sum_{s\in\mathcal{S}}E(C_i,B,s)\notag\\
&+\sum_{i=1}^{N}\sum_{s\in\mathcal{S}_{\mathrm{ZNE}}}F_sE(C_i,B,s)\Big).
\end{align}
The key observation is that replacing grid ordering with bandit ordering does not remove the inner loop over $\mathcal{S}$. Therefore the asymptotic candidate-evaluation count remains $O(NC)$.

\subsection{Pareto-Front Complexity}

The Pareto implementation normalizes metric values and groups duplicate objective vectors before maintaining an incremental non-dominated set. In the worst case, when most points are mutually non-dominated, multi-objective Pareto filtering still has quadratic behavior in the number of unique metric vectors. For the default candidate set of at most 23 strategies, this cost is negligible compared with transpilation and execution.

\subsection{Cache Semantics}

The compiled-circuit cache key is based on backend name, circuit fingerprint, and strategy JSON. This is sufficient for local reuse under stable software and backend conditions. It is not a complete archival provenance key because it does not encode all context that could affect compilation, such as Qiskit package version, live calibration timestamp, pass-manager implementation version, and profiling flag. A provenance-grade cache key would also include environment metadata and backend calibration identifiers.

\section{Evaluation Methodology}

The repository is evaluated here as a reproducible workflow artifact. The inspected release contains source code, tests, a small built-in circuit dataset, command-line workflows, and artifact writers; it does not contain an archived benchmark matrix over large circuit families or live quantum processors. The evaluation methodology therefore separates evidence that is available from the release itself from measurements that require running the package in a controlled environment. This avoids attributing performance improvements to \qb{} without executed data.

\subsection{Experimental Scope}

The evaluation is organized around three scopes.

\paragraph{Static implementation analysis.}
This scope inspects the implemented strategy space, objective, candidate ordering, Pareto filtering, caching contract, mitigation hooks, and report-generation path. It establishes properties that follow from the source code independently of backend noise or runtime availability.

\paragraph{Compile-only reproducibility.}
This scope compiles every circuit-strategy pair against a declared backend target and records structural metrics of the compiled circuit. It is sufficient for studying depth, circuit size, two-qubit-gate count, compile time, and the repository's estimated-error proxy. It does not establish execution fidelity.

\paragraph{Execution-backed validation.}
This scope requires a backend capable of returning shot counts. For simulator studies, the workflow can use Aer-compatible execution; for hardware studies, the backend must be supplied by the user and the resulting data must be archived with software versions, calibration context, shots, seeds, and backend identifiers. Qiskit Aer documents simulator-backed execution, while Qiskit Runtime exposes primitive options for resilience, twirling, and execution control when Runtime primitives are used directly \cite{QiskitAerSimulator2026,QiskitRuntimeOptions2026}.

\subsection{Datasets and Backends}

Let
\begin{equation}
\mathcal{D}=\{C_1,\ldots,C_N\}
\end{equation}
be the circuit dataset, where each $C_i$ is loaded from the dataset index and its associated QPY or QASM artifact. The built-in dataset provides only a minimal smoke-test set. Larger empirical studies require external workloads, for example routing-sensitive circuits, variational ansatz circuits, QFT-like circuits, QAOA-style circuits, or application circuits with nontrivial two-qubit interaction graphs. The workflow itself does not prescribe such a benchmark corpus; it only supplies the dataset container and execution path.

Backends are treated as explicit experimental inputs. For compile-only runs, a fake backend or backend-like target is enough to exercise Qiskit's preset pass-manager path. Qiskit preset pass managers transform an input circuit through staged transpilation passes and are configured by optimization level, backend target, layout method, routing method, translation method, and related options \cite{QiskitPassManagers2026,JavadiAbhari2024Qiskit}. For execution runs, the backend must support a \texttt{run} interface or be wrapped by an Aer simulator path when available.

\subsection{Reproducible Execution Protocol}

Listing~\ref{lst:protocol} gives a minimal local protocol. It is a reproducibility recipe for producing artifacts from the released package; the manuscript does not report numerical results from an unexecuted run.

\begin{lstlisting}[style=terminal,caption={Minimal local protocol for regenerating workflow artifacts.},label={lst:protocol}]
python -m venv .venv
source .venv/bin/activate
pip install -e ".[all]"

python -m qbalance dataset examples --out ./circuits --overwrite

python -m qbalance adjust ./circuits \
  --backend fake:generic:5 \
  --out ./balanced \
  --search bandit \
  --pareto \
  --max-candidates 24 \
  --overwrite

python -m qbalance matrix ./circuits \
  --backend fake:generic:5 \
  --backend fake:generic:10 \
  --out ./matrix.json \
  --execute \
  --shots 1024

python -m qbalance report ./matrix.json --out ./report --html
\end{lstlisting}

A controlled empirical run records the package version, Qiskit version, optional dependency versions, backend specification, seeds, shot count, candidate list, objective weights, and generated JSON artifacts. Without these fields, repeated runs can be difficult to interpret because transpilation behavior and backend calibration data can change over time.

\subsection{Evaluation Matrix}

Table~\ref{tab:evaluation} summarizes experiments that the workflow can run without changing the public API. The table does not imply that the measurements have already been performed; it defines the observable quantities required for a reproducible study.

\begin{table}[H]
\caption{Evaluation matrix for a reproducible \qb{} study.}
\label{tab:evaluation}
\centering
\begin{tabularx}{\textwidth}{>{\raggedright\arraybackslash}p{1.1cm}>{\raggedright\arraybackslash}p{2.5cm}>{\raggedright\arraybackslash}X>{\raggedright\arraybackslash}p{2.6cm}}
\toprule
ID & Scope & Controlled condition & Recorded quantities \\
\midrule
E1 & Compile-only selection & Fixed dataset, backend specification, objective weights, and candidate limit; compare grid and bandit ordering & Selected strategy, depth, size, two-qubit-gate count, compile time, estimated-error proxy \\
E2 & Pareto selection & Repeat E1 with Pareto filtering disabled and enabled & Selected strategy changes, Pareto-front membership, objective score \\
E3 & Simulator execution & Aer-compatible execution with fixed shots and seeds & Raw counts, count entropy, most probable bit string, execution failures if any \\
E4 & Readout mitigation path & Execution with M3-enabled candidates and declared calibration-shot setting & Mitigated probability summaries or captured mitigation errors \\
E5 & ZNE helper path & Execution with declared scale factors and polynomial degree & Folded-circuit factors, parity expectation, extrapolated scalar, pseudo-distribution summary \\
E6 & Cutting hook & Candidate with cutting enabled and max-subcircuit size specified & Cut success or failure, transformed circuit metadata when available \\
E7 & Reporting path & Matrix JSON rendered through Markdown or HTML report command & Report files, aggregate tables, serialized input matrix \\
\bottomrule
\end{tabularx}
\end{table}

\subsection{Metrics}

For each circuit $C_i$, let $\widetilde{C}_{i,\mathrm{base}}$ be the compiled baseline circuit and let $\widetilde{C}_{i,s_i^*}$ be the compiled circuit produced by the selected strategy. Compile-level comparisons are reported as ratios only when the corresponding baseline denominator is nonzero:
\begin{align}
R_{\mathrm{depth}}(i) &=
\frac{\mathrm{depth}(\widetilde{C}_{i,s_i^*})}
     {\mathrm{depth}(\widetilde{C}_{i,\mathrm{base}})},\\
R_{2q}(i) &=
\frac{\mathrm{2q}(\widetilde{C}_{i,s_i^*})}
     {\mathrm{2q}(\widetilde{C}_{i,\mathrm{base}})},\\
\Delta \widehat{E}(i) &=
\widehat{E}(\widetilde{C}_{i,s_i^*})-
\widehat{E}(\widetilde{C}_{i,\mathrm{base}}).
\end{align}
For shot-count data $c(x)$ over bit strings $x$, with $p(x)=c(x)/\sum_y c(y)$, the workflow records
\begin{align}
H(p) &= -\sum_x p(x)\log_2 p(x),\\
p_{\max} &= \max_x p(x).
\end{align}
These execution summaries describe the observed distribution but are not, by themselves, fidelity measures. Observable-level accuracy studies require an explicitly declared observable, an estimator, and uncertainty analysis. This distinction is important for error mitigation: readout mitigation and zero-noise extrapolation are normally evaluated through corrected distributions or expectation-value estimators rather than through the largest observed bit-string probability \cite{Nation2021,Temme2017,Cai2023QEM}.

\section{Implementation-Level Findings}

\subsection{Workflow Scope}

\qb{} is an orchestration layer over established compilation and execution mechanisms. Compilation is delegated to Qiskit preset pass managers; execution is delegated to a backend or to Aer-compatible execution when a backend does not directly expose a runnable interface; M3 mitigation is delegated to the external matrix-free mitigation package. The implemented contribution is the organization of these mechanisms into a dataset-level selection workflow with serializable strategy specifications, objective scoring, Pareto filtering, caching, diagnostics, and report generation \cite{qbalanceRepo,QiskitPassManagers2026,MthreeDocs2026}.

This scope is narrower than a replacement compiler or a new physical error-mitigation method. The release supports reproducible comparison of workflow choices, but it does not by itself establish new routing optimality, new error-suppression physics, or hardware-level accuracy gains.

\subsection{Candidate Space}

The default candidate generator constructs four groups of strategies: compilation sweeps over Qiskit optimization levels and layout/routing choices, suppression variants, mitigation variants, and one cutting variant. The resulting default set contains at most 23 unique strategies. This finite and compact candidate set makes exhaustive candidate evaluation practical on small datasets. It also explains why the current bandit mode affects candidate ordering rather than reducing the number of candidates evaluated.

\subsection{Bandit Ordering}

The bandit module uses a Bayesian linear surrogate over a fixed feature vector derived from \texttt{StrategySpec}. The posterior update is mathematically consistent with ridge-regularized Bayesian linear regression, and proposal uses a Thompson-style coefficient sample. However, the feature vector contains strategy attributes only; it does not include circuit descriptors such as qubit count, initial depth, gate histogram, or interaction-graph statistics. The learned preference is therefore global across evaluated strategies rather than contextualized to individual circuits.

The current evaluation loop also visits every candidate in the constructed order. As implemented, bandit mode can change the sequence in which strategies are evaluated, but it does not reduce the compile or execution budget relative to grid mode. A budgeted variant would require an explicit evaluation cap or a stopping rule.

\subsection{Error Proxy}

The repository's estimated-error metric is a product-of-survival heuristic. If $e_j$ denotes the selected per-operation error contribution, the estimator has the form
\begin{equation}
\widehat{E}=1-\prod_j(1-e_j).
\end{equation}
For small $e_j$, the first-order expansion is
\begin{equation}
\widehat{E}=\sum_j e_j + O(e^2).
\end{equation}
This makes the proxy monotone with respect to additional error-bearing operations under the assumptions encoded by the implementation. It does not model coherent errors, correlated errors, crosstalk, calibration drift, or observable-specific bias. It is therefore suitable as a lightweight ranking feature, not as a physical fidelity estimator.

\subsection{Mitigation Hooks}

The M3 path calls an external matrix-free measurement-mitigation implementation. M3 is designed to mitigate measurement assignment errors without explicitly forming the full assignment matrix, working instead in a reduced subspace associated with observed noisy bit strings \cite{Nation2021,MthreeDocs2026}. In \qb{}, this path returns corrected probability summaries when the optional dependency and backend calibration path are available; errors are recorded in metrics rather than terminating the full workflow.

The ZNE helper is more limited. It executes folded circuits, converts counts to a parity expectation, extrapolates that scalar to zero noise, and adjusts even and odd probability mass in a base distribution. This is not a general observable-level ZNE framework. Foundational ZNE estimates zero-noise observables by evaluating a chosen observable at multiple amplified noise levels and extrapolating the observable value to the zero-noise limit \cite{Temme2017,Endo2018,Cai2023QEM}. Accordingly, the repository's helper is treated here as an experimental counts-based utility rather than as a complete observable-level ZNE implementation.

\subsection{Circuit-Cutting Hook}

The cutting path is best interpreted as a preprocessing hook. Qiskit addon cutting describes circuit cutting as a decomposition of a larger circuit into subexperiments followed by classical reconstruction, with a sampling overhead determined by the cuts \cite{QiskitAddonCutting2026}. The inspected workflow invokes cut finding before compilation, but it does not expose a complete subexperiment execution and reconstruction pipeline as a first-class selection objective. Consequently, the release supports cut-aware experimentation at the transformation stage, not a complete circuit-knitting benchmark.

\section{Discussion}

\subsection{Interpretation of the Selection Rule}

The central operation in \qb{} is a per-circuit selection map
\begin{equation}
(C_i,B,\mathcal{S},J_w) \longmapsto s_i^*,
\end{equation}
where $C_i$ is an input circuit, $B$ is the backend specification, $\mathcal{S}$ is a finite candidate strategy set, $J_w$ is the weighted objective, and $s_i^*$ is the selected strategy. This formulation captures the implemented behavior without treating any candidate as universally best. The selected strategy is conditional on the backend target, objective weights, available optional dependencies, execution setting, and whether Pareto filtering is enabled.

The formulation also clarifies the role of the default objective. The score combines structural metrics and a lightweight error proxy. It is useful for comparing candidates inside the same workflow, but it is not a substitute for an application-level observable, a fidelity benchmark, or a hardware-calibrated error model. This distinction is essential when interpreting selected strategies from compile-only runs.

\subsection{Use in Reproducible Quantum-Software Experiments}

The package is most directly applicable to studies where the experimental object is the workflow choice itself: the effect of optimization level, routing method, layout heuristic, suppression toggle, mitigation toggle, or Pareto filtering on a dataset of circuits. In that setting, the value of \qb{} is the explicit representation of strategies and the preservation of outputs in machine-readable artifacts. The same structure can also support regression tests for quantum-software pipelines, smoke tests for optional mitigation dependencies, and comparative matrix reports across backend specifications.

The current implementation is less suited to claims about physical improvement unless additional measurements are added. A hardware study would require archived backend identifiers, calibration context, shot counts, seeds, raw counts, observable definitions, and uncertainty estimates. Without those records, the output of the workflow remains a strategy-selection result rather than evidence of improved quantum accuracy.

\section{Limitations}

The analysis identifies eight implementation-level limitations.

\begin{enumerate}
\item \textbf{Benchmark coverage.} The repository contains a minimal built-in dataset and tests, but no archived large-scale benchmark matrix over real devices or broad circuit families.
\item \textbf{Search budget.} Bandit mode ranks candidates adaptively, but the workflow evaluates the full candidate set in the inspected implementation.
\item \textbf{Circuit context.} The bandit surrogate uses strategy features only. Circuit properties such as qubit count, two-qubit interaction graph, initial depth, and gate composition are not part of the model.
\item \textbf{Layout heuristic.} The custom noise-aware layout ranks qubits with a greedy heuristic and does not solve a topology-aware placement problem over pairwise gate errors and routing distances.
\item \textbf{Error model.} The estimated-error metric is a product-of-survival proxy. It does not model coherent errors, correlated errors, crosstalk, calibration drift, or observable-specific bias.
\item \textbf{ZNE helper.} The counts-based ZNE path extrapolates a parity expectation and constructs a pseudo-distribution. It is not a general ZNE implementation for arbitrary observables.
\item \textbf{Circuit cutting.} The cutting path invokes cut finding as a transformation step, but complete subexperiment generation, execution, reconstruction, and sampling-overhead-aware scoring are not integrated as first-class workflow outputs.
\item \textbf{Runtime integration.} Runtime-style options can be constructed, but the main workflow does not execute Qiskit Runtime \texttt{EstimatorV2} or \texttt{SamplerV2} jobs.
\end{enumerate}

\section{Threats to Validity}

\subsection{Internal Validity}

The conclusions are derived from the current implementations used by the workflow. The analysis does not include exhaustive execution of every optional path on live hardware. Optional dependencies can change behavior across releases, and backend APIs can differ between simulators, fake backends, and managed hardware services.

\subsection{External Validity}

The built-in circuits are small and serve as smoke tests rather than as a representative benchmark suite. Structural conclusions about the workflow generalize to other datasets because they follow from the implementation. Performance conclusions do not generalize without running the workflow on larger datasets and explicitly recording the experimental environment.

\subsection{Construct Validity}

Depth, two-qubit-gate count, compile time, and estimated error are proxy variables. They can explain structural trade-offs introduced by compilation and routing, but they do not fully characterize task success. For an algorithmic workload, the relevant quantity is usually an observable, distributional distance, success probability, or estimator variance tied to the application.

\section{Conclusion}

\qb{} provides a dataset-level workflow for comparing quantum compilation, suppression, and mitigation strategies under a common objective. Its implementation combines dataset management, backend resolution, immutable strategy specifications, Qiskit preset-pass-manager compilation, optional execution and mitigation hooks, scalar and Pareto selection, caching, diagnostics, and report generation. The resulting artifact is useful because it turns ad hoc tuning choices into explicit and reproducible workflow decisions.

The strongest supported conclusion is architectural rather than empirical. The package implements a coherent strategy-selection workflow, while compilation, simulation, readout mitigation, runtime resilience options, and circuit cutting remain grounded in external tools and literature. The inspected release does not establish hardware-level accuracy improvement, optimal routing, or a complete circuit-knitting pipeline. Those claims require additional experiments with archived datasets, backends, observables, shot counts, and uncertainty estimates.

\section*{Data and Code Availability}

The analyzed implementation is available in the public \pkg{} repository \cite{qbalanceRepo}. This manuscript does not introduce new experimental datasets or hardware-measurement results. The built-in examples are intended for smoke-test reproduction and workflow validation.

\section*{Acknowledgments}

The author thanks the maintainers of Qiskit, Qiskit Aer, Qiskit Runtime, M3, and Qiskit addon cutting for making the underlying quantum software stack and documentation available to the community.

\bibliographystyle{splncs04}
\bibliography{refs}

\end{document}